\documentclass[10pt,aps,prd,twocolumn,superscriptaddress,floatfix,amsmath,amssymb,amsfonts,longbibliography,nofootinbib]{revtex4-2}

\usepackage[english]{babel}
\usepackage{graphicx}
\usepackage{dcolumn}
\usepackage{bm}

\usepackage{physics}
\usepackage{amsmath,amssymb,mathtools}
\usepackage{booktabs}
\usepackage{tikzsymbols}
\usepackage[utf8]{inputenc}
\usepackage[T1]{fontenc}
\usepackage{array}
\usepackage{booktabs}

\usepackage{tikz}
\usetikzlibrary{arrows.meta, positioning}
\usepackage{lipsum}
\usepackage{hyperref}

\usepackage{pgfplots}
\pgfplotsset{compat=1.18}

\hypersetup{
    colorlinks=true,
   linkcolor=blue,
    filecolor=red,      
    urlcolor=cyan,
}

\begin{document}

\title{Spin-Induced Fractal Time-Crystal-Like Dynamics and Non-Markovian Memory \protect\\ in the Bateman Dual Oscillator}




\author{Partha Nandi}
\email{pnandi@sun.ac.za}
\affiliation{Department of Physics, University of Stellenbosch, Stellenbosch 7600, South Africa}
\affiliation{National Institute of Theoretical and Computational Sciences (NITheCS), Stellenbosch 7604, South Africa}

\author{Giuseppe Vitiello}
\email{givitiello@unisa.it}
\affiliation{Physics Department ``E.R. Caianiello'', University of Salerno,
Via Giovanni Paolo II, 132, 84084 Fisciano (Salerno), Italy}

\begin{abstract}
Can a closed quantum system generate  
time-crystal-like nonequilibrium behavior, self-similar scaling structures, and non-Markovian memory without external driving or coupling to a macroscopic environment? We address this question within the quantum Bateman oscillator formulated in a nonrelativistic $(2+1)$-dimensional phase-space noncommutative framework generated by spin-induced spatial deformation. The resulting doubled quantum dynamics is governed by a time-independent Hermitian Hamiltonian and exhibits an underlying $SU(1,1)$ structure with amplified and damped collective modes. We show that these modes satisfy an exact discrete scaling covariance, leading to self-similar temporal evolution without external driving. Upon tracing over one oscillator sector, the reduced dynamics becomes intrinsically non-Markovian and is governed by a history-dependent memory kernel. The same scaling structure admits a geometric representation in terms of logarithmic-spiral trajectories associated with the amplified and damped branches of the Bateman system. 
Because the mechanism relies on nonequilibrium reduced dynamics rather than equilibrium expectation values of local observables, it lies outside the assumptions underlying conventional no-go theorems for equilibrium time crystals. Our results identify spin as the common physical origin of the amplified and damped Bateman dynamics, self-similar scaling periodicity, logarithmic-spiral structures, and non-Markovian memory, also suggesting a natural extension of the mechanism to relativistic anyonic systems.

\end{abstract}

\maketitle

\section{Introduction}

The possibility that a physical system may display persistent
periodic motion in time despite being governed by a time–independent
Hamiltonian has attracted considerable attention in recent years \cite{PhysRevLett.109.160402}.
This phenomenon was originally proposed by Wilczek in the context
of ``time crystals,'' where the ground state of a quantum system
would spontaneously break continuous time–translation symmetry
\cite{Wilczek:2012jt, Nandi:2024cev}. Although subsequent analyses showed that
such spontaneous symmetry breaking cannot occur in equilibrium
ground states of time–independent many–body Hamiltonians \cite{Watanabe:2014hea},
the concept has stimulated extensive research into systems
exhibiting time–periodic behavior under more general conditions \cite{Bruno:2013}.

A particularly successful realization is provided by
\emph{Floquet time crystals} \cite{Else:2016}, where periodic driving
leads to a discrete breaking of time–translation symmetry
in the response of many–body systems \cite{
Khemani:2016, Yao:2017}. In these systems, the Hamiltonian
itself is explicitly time dependent, and the periodicity
results from the interplay between external driving,
interactions, and many–body localization. Experimental realizations have been reported in several
platforms, most notably in trapped-ion systems \cite{Zhang:2017}
and disordered spin ensembles \cite{Choi:2017}. More recently, fractional time crystals have been proposed,
where the emergent temporal periodicity may correspond to a
rational multiple of the driving period, further enriching
the possible forms of temporal ordering in periodically
driven quantum systems \cite{PhysRevA.99.033626}.

 Beyond driven systems, time--crystal--like behavior
may also arise through alternative nonequilibrium mechanisms.
In particular, gravitational interactions can induce persistent
oscillatory behavior in reduced subsystems through coherent
exchange and memory effects, without relying on conventional
Floquet driving or equilibrium spontaneous symmetry breaking
\cite{Dutta:2025bge, Dutta:2025ouy}.


It is therefore natural to ask whether persistent temporal oscillations
can arise in systems with time–independent Hamiltonians through
alternative mechanisms. One possibility involves noncommutative
phase–space structures \cite{PhysRevD.74.045015}, where the canonical coordinates or momenta
obey deformed commutation relations. Such deformations can induce
couplings between degrees of freedom even in otherwise
simple systems. In particular, it has been shown that noncommutative
harmonic oscillators can exhibit periodic dynamics in subsystem
observables reminiscent of time–crystal behavior \cite{Bernardini:2022vec}.

Modified phase--space structures naturally arise in the
dynamics of dissipative systems. A paradigmatic example
is the Bateman dual oscillator, which provides a
Hamiltonian embedding of a damped system through the
introduction of an additional amplified degree of
freedom \cite{Bateman1931}. In its first-order
formulation, the Bateman system contains a
Chern--Simons--like term in momentum space which,
through Dirac constraint analysis, induces a
noncanonical symplectic structure. Consequently, the
configuration-space coordinates acquire nontrivial
Dirac brackets, generating an effective classical
noncommutative phase space. In this way, dissipation
itself can be understood as a source of
noncommutativity \cite{PhysRevA.97.062110},
establishing a direct connection between dissipative
dynamics and noncommutative geometry \cite{Banerjee:2001zi}. Related
open-system perspectives \cite{Breuer:2007juk} in which quantum spacetime structures \cite{Doplicher:1994zv, PhysRevD.108.086003} effectively induce
reduced subsystem dynamics and dissipative behavior
have also recently been explored in
Ref.~\cite{Nandi:2025qyj}. Such structures are further
closely related to the generalized (extended)
Newton--Hooke algebra in $(2+1)$ dimensions,
$\overline{NH}_{3}$ \cite{Pal:2024btm}.

Closely related oscillatory
regimes have also been identified in PT-symmetric realizations of the
Bateman system, where balanced gain–loss dynamics leads to bounded and
periodic behavior of observables \cite{Beetar:2023mfn}. 

More broadly, Chern--Simons--type structures and their noncommutative
extensions \cite{PhysRevLett.87.030402,Gangopadhyay:2014cea} have played a central role in the quantum Hall effect, where
noncommutative Chern--Simons theories provide effective descriptions
of electrons in the lowest Landau level and successfully account for
the observed filling fractions
\cite{Bellissard1994,Morariu2001,Hellerman2001}. Related noncommutative
and constrained phase-space descriptions have also recently been developed
for lowest-Landau-level vortex systems, Tkachenko collective modes,
and projected Landau-level dynamics, where the effective degrees of freedom
naturally acquire noncommutative geometric structures and reduced
phase-space descriptions
\cite{Manoj:2025knl,PhysRevResearch.6.L012040,Mandal:2025ofl}. These connections further
highlight that noncommutative structures are closely related to physically
realizable systems \cite{Nandi:2023xmo}.

The interplay between dissipation, noncanonical symplectic structures,
and quantization has been extensively investigated,
notably by Banerjee \textit{et al.} \cite{Banerjee:2001zi,Banerjee:2001yc}.
It has also been shown that the
dissipative dynamics of open systems can be consistently embedded within
quantum field theory by ``closing'' the system through a doubling of
degrees of freedom \cite{Celeghini:1991yv}, thereby introducing an amplified counterpart,
characteristic of the Bateman construction \cite{Bateman1931,Feshbach1977}, also in connection with noncommutative geometry \cite{Sivasubramanian:2003xy}, oscillatory models in general relativity \cite{Pashaev} and fractal self-similarity \cite{Vitiello2012}.

Motivated by 't Hooft's proposal \cite{Hooft:1999}, it has further been shown \cite{Blasone:2000ew} that classical information loss, under suitable boundary conditions, can be effectively described in terms of a quantization scheme. Moreover, it has been clarified that dissipative systems admit a formulation in terms of gauge-theoretic structures defined on a pseudo-Euclidean plane, thereby uncovering a profound connection between dissipative dynamics and topologically massive Chern-Simons  gauge theories \cite{Asorey:1993ft}, also with application to the dynamics of Bloch electrons in a solid \cite{Blasone:1996yh}.

In this work, building on these developments, we investigate the interplay between dissipation, noncommutativity, self-similar scaling structures, and memory effects in quantum systems. Starting from a noncanonical symplectic structure that induces a noncommutative phase space and gives rise to the Bateman dual oscillator, we construct a canonical representation and analyze the resulting quantum dynamics.

We show that the quantized doubled Bateman system possesses an underlying $\mathrm{SU}(1,1)$ dynamical structure with amplified and damped collective modes whose evolution generates an exact self-similar scaling periodicity. The corresponding collective observables satisfy discrete scaling relations under finite time evolution, revealing a temporal self-similarity that is fundamentally distinct from the periodic behavior encountered in conventional Floquet systems. The same dynamics admits a natural geometric realization in terms of logarithmic-spiral trajectories and fractal-like self-similar structures.

We further demonstrate that tracing over one oscillator sector leads to an effective reduced description governed by a non-Markovian evolution equation containing a history-dependent memory kernel \cite{Rodriguez-Rosario:2008zyc, Nandi:2026muw}. Although the collective observables responsible for the scaling periodicity become inaccessible after the partial trace, the correlations associated with the amplified and damped branches remain encoded in the full density matrix and continue to influence the observable subsystem through the memory kernel. The resulting reduced dynamics therefore retains an indirect imprint of the hidden self-similar structure of the underlying doubled theory.

Because the mechanism relies on nonequilibrium reduced dynamics rather than equilibrium expectation values of local observables, it lies outside the assumptions underlying conventional no-go theorems for equilibrium time crystals \cite{Watanabe:2014hea}. Related nonequilibrium and dissipative time-crystal scenarios have also been investigated in open quantum systems and Liouvillian frameworks \cite{PRXQuantum.5.030325}. The present construction therefore provides a unified framework in which amplified and damped Bateman dynamics, logarithmic-spiral self-similarity, discrete scaling periodicity, and non-Markovian memory emerge as interconnected manifestations of the same underlying quantum dynamics.

The paper is organized as follows. In Sec.~II we review the Bateman oscillator and its noncommutative Dirac-bracket structure. In Sec.~III we introduce a canonical Bopp--shift representation, rewrite the Hamiltonian in light-cone variables, and proceed to its quantization, obtaining the corresponding two-mode Hamiltonian and solving the Heisenberg equations of motion. In Sec.~IV we analyze the amplified and damped collective modes and establish the resulting self-similar scaling periodicity of the doubled dynamics. In Sec.~V we investigate logarithmic-spiral and Koch-type self-similar structures and discuss their relation to the underlying scaling dynamics and non-Markovian reduced evolution. Sec.~VI contains concluding remarks and possible extensions of the present framework. Additional technical details are presented in Appendices A and B.

\section{Bateman Oscillator from Symplectic Deformation Induced by Fractional Spin in 2+1 Dimensions}

Planar dynamical systems possess a distinctive symmetry
structure. It is well known that the Galilei group in
two spatial dimensions admits a two-dimensional central
extension characterized by the mass $m$ and an
additional parameter $s$
\cite{LevyLeblond1972,bose1995galilean,Papageorgiou:2009zc}.
In particular, the Galilean boost generators
$K_i$ satisfy the nontrivial algebra
\begin{equation}\label{1a}
\{K_i,K_j\}
=
s\,\epsilon_{ij},
\qquad i,j=1,2.
\end{equation}

Let us consider the classical phase–space realization of the boosts in terms
of coordinates $Y_i$ and momenta $\pi_i$,
\begin{equation}
K_i = m Y_i - t\,\pi_i .
\end{equation}
If the phase space is canonical,
\begin{equation}
\{Y_i,Y_j\}=0, \qquad
\{Y_i,\pi_j\}=\delta_{ij}, \qquad
\{\pi_i,\pi_j\}=0 ,
\end{equation}
one immediately finds
\begin{equation}
\{K_i,K_j\}=0 ,
\end{equation}
which 
does not reproduce the 
commutation relation (\ref{1a}).
Thus, the realization of the planar Galilei algebra with the second central charge requires a modification of the symplectic structure
of the phase space.

As shown by Jackiw and Nair~\cite{Jackiw:2000tz}, planar particles with
arbitrary spin \cite{PhysRevD.106.L121503} are described by a modified phase--space symplectic
two--form
\begin{equation}
\Omega =
d\pi_i \wedge dY_i
+
\frac{s}{2m^2}\epsilon_{ij}\,
d\pi_i \wedge d\pi_j ,
\end{equation}
where the additional $d\pi \wedge d\pi$ term encodes the exotic
central extension of the planar Galilei group and induces a
nontrivial symplectic structure for the spatial coordinates.

This symplectic structure can be generated from the symplectic
one--form
\begin{equation}
\Theta =
\pi_i\, dY_i
+
\frac{s}{2m^{2}}\epsilon_{ij}\pi_i\, d\pi_j .
\end{equation}
 The corresponding first--order action
then takes the form
\begin{equation}
S[\pi_{i}(t),Y_{i}(t)] = \int dt
\left(
\pi_i \dot Y_i
+ \frac{s}{2m^{2}}\epsilon_{ij}\pi_i \dot\pi_j
- H(Y,\pi)
\right),
\end{equation}
which provides a convenient starting point for the description of
planar systems with modified symplectic structure.\\

The variables $\pi_i$ are introduced as independent phase--space
coordinates in the first--order formulation of the indirect planar
oscillator. This formulation allows the symplectic structure of the
phase space to be modified by the addition of a Chern--Simons--like
term in momentum space. Although the canonical momenta conjugate to
$Y_i$ eventually coincide with $\pi_i$, these variables cannot be
eliminated at the level of the Lagrangian because they encode the
deformation of the symplectic two-form that governs the modified
dynamics associated with arbitrary spin in planar nonrelativistic
systems.

In the present work we consider the indirect representation of a
planar oscillator \cite{Santilli1984}, whose Hamiltonian is
\begin{equation}
H =
\frac{1}{m}\pi_1 \pi_2
+
m\Omega^2 Y_1 Y_2 .
\label{1}
\end{equation}
The corresponding Lagrangian becomes
\begin{equation}
L =
\pi_i \dot Y_i
+ \frac{s}{2m^{2}}\epsilon_{ij}\pi_i \dot\pi_j
- \frac{1}{m}\pi_1 \pi_2
- m\Omega^2 Y_1 Y_2 .
\label{fk}
\end{equation}

The canonical momenta conjugate to $Y_i$ are
\begin{equation}
P_{Y_i} = \frac{\partial L}{\partial \dot Y_i} = \pi_i ,
\end{equation}
while those conjugate to $\pi_i$ are
\begin{equation}
P_{\pi_1} = -\frac{s}{2m^{2}}\pi_2 ,
\qquad
P_{\pi_2} = \frac{s}{2m^{2}}\pi_1 .
\end{equation}

These relations give rise to the primary constraints
\begin{equation}
\phi_1 = P_{\pi_1} + \frac{s}{2m^{2}}\pi_2 \approx 0,
\qquad
\phi_2 = P_{\pi_2} - \frac{s}{2m^{2}}\pi_1 \approx 0 .
\end{equation}

The constraints $\phi_1$ and $\phi_2$ form a set of second--class
constraints in the sense of Dirac \cite{Dirac1950, Dirac1959, Dirac1964}, since their Poisson bracket
matrix is non--vanishing. Consequently the system possesses no
gauge freedom and no Lagrange multipliers are required to enforce
the constraints. In such cases the appropriate framework is to
replace the canonical Poisson brackets by Dirac brackets, defined
for two phase--space functions $A$ and $B$ as
\begin{equation}
\{A,B\}_{DB}
=
\{A,B\}
-
\{A,\phi_\alpha\}
C^{-1}_{\alpha\beta}
\{\phi_\beta,B\},
\end{equation}
where $C_{\alpha\beta}=\{\phi_\alpha,\phi_\beta\}$ is the constraint
matrix. The use of Dirac brackets allows the second--class
constraints to be imposed strongly, effectively eliminating the
redundant degrees of freedom from the phase space.

Following Dirac's procedure for constrained systems,
the effective Dirac brackets become
\begin{align}
\{Y_i,Y_j\}_{DB} &= \frac{s}{m^{2}}\,\epsilon_{ij}, \\
\{Y_i,\pi_j\}_{DB} &= \delta_{ij}, \\
\{\pi_i,\pi_j\}_{DB} &= 0 .
\label{align}
\end{align}
The deformation term thus generates an effective
classical noncommutative phase-space structure through
the nontrivial brackets between the coordinates.

Employing the Dirac brackets together with the
Hamiltonian equations generated by the Hamiltonian
(\ref{1}), one obtains the following equations of
motion for the phase-space variables:
\begin{equation}
\ddot{Y}_1 - \gamma \dot{Y}_1 + \Omega^2 Y_1 = 0,
\qquad
\ddot{Y}_2 + \gamma \dot{Y}_2 + \Omega^2 Y_2 = 0 ,
\end{equation}
where 

\begin{equation}
\gamma = \frac{\Omega^2 s}{m}.
\end{equation}
The dissipative dynamics of the Bateman oscillator therefore results
entirely from the modified symplectic structure of the phase space.
In this sense the present model establishes a direct correspondence
between noncommutativity and dissipation (Bateman system),
\[
\text{Spin-induced spatial deformation}
\,\longleftrightarrow \,
\text{dissipation} .
\]

In the next section, we quantize the system and analyze its operator formulation.

\section{Light-cone quantization of the Bateman system}

Upon quantization, the Dirac brackets among the phase–space variables 
(\ref{align}) are elevated to commutators via the correspondence principle, leading
to the algebra
\begin{equation} \label{cy}
[\hat{Y}_i,\hat{Y}_j]= i\theta\,\epsilon_{ij}\mathbb{I}, \quad
[\hat{Y}_i,\hat{\pi}_j] = i\hbar\delta_{ij}\mathbb{I}, \quad
[\hat{\pi}_i,\hat{\pi}_j]= 0 ,
\end{equation}
where
\begin{equation}
\theta \sim \hbar s/m^{2}.\end{equation}

To perform canonical quantization, it is convenient to introduce a set of
canonical phase–space variables $(X_i,P_i)$, $i = 1,2$, satisfying the standard
Heisenberg algebra
\begin{equation}
[\hat{X}_i,\hat{X}_j]=0,
\quad
[\hat{X}_i,\hat{P}_j]=i\hbar\delta_{ij}\mathbb{I},
\quad
[\hat{P}_i,\hat{P}_j]=0 .
\end{equation}

The noncommutative variables can be represented in terms of these
operators through the Bopp shift \cite{Chou:1993bb,Jackiw:2000tz, Nandi:2023xmo}
\begin{equation}
\hat{Y}_i = \hat{X}_i - \frac{\theta}{2\hbar}\epsilon_{ij}\hat{P}_j;
\qquad
\hat{\pi}_i = \hat{P}_i ,
\end{equation}
 so that the commutation relations (\ref{cy}) are satisfied.

It is important to clarify the role of the canonical variables
$(X_i,P_i)$ introduced through the Bopp-type shift. 
They do not satisfy
the same algebra as the original 
$(Y_i,\pi_i)$
and should therefore be regarded as an auxiliary representation used to
implement the noncommutative phase-space structure in a standard Hilbert
space. A similar realization of noncommutative coordinates in terms of
canonical variables appears in the symplectic description of anyons
\cite{Chou:1993bb}, where the nontrivial commutation relations arise from
spin degrees of freedom.

The physical observables of the system are defined in terms
of the noncommutative  $(Y_i,\pi_i)$, while the canonical variables
$(X_i,P_i)$ serve as a computational tool. Although the analysis is
performed in the canonical representation, the resulting dynamical
features—such as the time-periodic behavior of observables—are determined
by the spectral structure of the Hamiltonian and are therefore independent
of the specific choice of variables. Consequently, the periodicity
identified below reflects a genuine physical property of the system and
persists when observables are expressed in terms of the original
noncommutative variables. In particular, the observables considered in the
present work are constructed from the Hamiltonian and its associated 
operators, and therefore inherit this representation independence.

In this representation, the noncommutative deformation is effectively
shifted from the algebra to the Hamiltonian, where it manifests as
additional interaction terms encoding the underlying spin-induced
structure.

The effects of
noncommutativity become manifest at the level of the Hamiltonian written in terms of the canonical variables.

To make the oscillator structure more transparent, it is convenient to introduce the light-cone variables
\begin{equation}
\hat X_\pm
=
\frac{\hat X_1\pm \hat X_2}{\sqrt{2}},
\qquad
\hat P_\pm
=
\frac{\hat P_1\pm \hat P_2}{\sqrt{2}}.
\end{equation}

In terms of these variables, the Hamiltonian becomes 
\begin{equation}
\begin{aligned}
\hat H
=&\;
\frac12
\left(
m\Omega^2\hat X_+^2
+
A\hat P_+^2
\right)
-  \,
\frac12
\left(
m\Omega^2\hat X_-^2
+
A\hat P_-^2
\right)
\\[4pt]
&+
\Gamma
\left(
\hat X_+\hat P_-
+
\hat X_-\hat P_+
\right),
\end{aligned}
\end{equation}
where
\begin{equation}
\label{Gamma}
A=
\frac1m
-
\frac{m\Omega^2\theta^2}{4\hbar^2},
\qquad
\Gamma=
\frac{m\Omega^2\theta}{2\hbar}.
\end{equation}

Introducing the effective mass parameter
\begin{equation}
m_\theta
\equiv
\frac1A
=
\frac{m}
{
1-\dfrac{m^2\Omega^2\theta^2}{4\hbar^2}
},
\end{equation}
the quadratic part of the Hamiltonian may be written in the standard oscillator form
\begin{equation}
\frac12
\left(
m\Omega^2  X_\pm^2
+
A P_\pm^2
\right)
=
\frac{P_\pm^2}{2m_\theta}
+
\frac12 m_\theta \Omega_{0}^2 X_\pm^2 ,
\end{equation}
where the effective frequency is
\begin{equation}
\Omega_{0}^2
=
\frac{m}{m_\theta}\Omega^2
=
\Omega^2-\Gamma^2 ,
\end{equation}
thus ${m_\theta}\Omega_{0}^2 = m \Omega^2$.

Quantization is implemented by promoting the canonical variables to operators satisfying the standard canonical commutation relations
\begin{equation}
[\hat X_\pm,\hat P_\pm]
=
i\hbar\mathbb I,
\qquad
[\hat X_+,\hat P_-]
=
[\hat X_-,\hat P_+]
=
0.
\end{equation}

It is then natural to introduce ladder operators adapted to the effective mass $m_\theta$,
\begin{equation}
\label{ab}
\begin{aligned}
a
&=
\sqrt{\frac{m_\theta\Omega_{0}}{2\hbar}}\,
\hat X_+
+
\frac{i}{\sqrt{2m_\theta\Omega_{0}\hbar}}\,
\hat P_+ ,
\\[4pt]
b
&=
\sqrt{\frac{m_\theta\Omega_{0}}{2\hbar}}\,
\hat X_-
+
\frac{i}{\sqrt{2m_\theta\Omega_{0}\hbar}}\,
\hat P_- .
\end{aligned}
\end{equation}

These operators satisfy the canonical bosonic commutation relations
\begin{equation}
[a,a^\dagger]=1,
\qquad
[b,b^\dagger]=1,
\end{equation}
with all other commutators vanishing.

Hence the full Hamiltonian can be written as
\begin{equation}
\hat H
=\hbar\Omega_{0}
\left(
a^\dagger a-b^\dagger b
\right)
+
i\hbar\Gamma
\left(
a^\dagger b^\dagger-ab
\right)
.
\label{H_ab}
\end{equation}

The operators $a$ and $b$ annihilate the vacuum states of the two-mode Fock space, $a|0\rangle_a = 0$, $b|0\rangle_b = 0$,
\begin{equation}
\mathcal{H}
=
\mathrm{span}
\left\{
|n,m\rangle
=
\frac{(a^\dagger)^n}{\sqrt{n!}}\,|0\rangle_a
\otimes
\frac{(b^\dagger)^m}{\sqrt{m!}}\,|0\rangle_b
\;\Big|\;
n,m\in\mathbb{N}_0
\right\},
\label{Hi}
\end{equation}
%
%

The Heisenberg equations of motion,
\begin{equation}
\dot O(t)=\frac{i}{\hbar}[\hat{H},O(t)],
\end{equation}
yield the coupled first-order differential equations 

%
\begin{equation}
\dot a
=
-i\Omega_0 a
+
\Gamma b^\dagger,
\qquad
\dot b^\dagger
=
-i\Omega_0 b^\dagger
+
\Gamma a .
\label{heisab}
\end{equation}
Introducing the rotating-frame operators
\(
\tilde a(t)=e^{i\Omega_0 t}a(t)
\)
and
\(
\tilde b^\dagger(t)=e^{i\Omega_0 t}b^\dagger(t),
\)
the coupled equations can be decoupled and solved exactly. One obtains
\begin{equation}
\tilde b^\dagger(t)
=
\cosh(\Gamma t)\,\tilde b^\dagger(0)
+
\sinh(\Gamma t)\,\tilde a(0).
\end{equation}
Transforming back to the original operators yields
\begin{equation}
a(t)
=
u(t)\,a(0)
+
v(t)\,b^\dagger(0),
\label{bogo}
\end{equation}
where
\begin{equation}
u(t)
=
e^{-i\Omega_0 t}\cosh(\Gamma t),
\qquad
v(t)
=
e^{-i\Omega_0 t}\sinh(\Gamma t).
\end{equation}
Similarly, the evolution of \(b(t)\) is given by
\begin{equation}
b(t)
=
u(t)^{*}\,b(0)
+
v(t)^{*}\,a^\dagger(0).
\label{g}
\end{equation}

The resulting time evolution therefore takes the form of an $SU(1,1)$ Bogoliubov transformation. Since the Bogoliubov coefficients satisfy $|u(t)|^{2}-|v(t)|^{2}=1$, the transformation is canonical, preserving the bosonic canonical commutation relations, $[a(t),a^\dagger(t)]=[b(t),b^\dagger(t)]=1$, with all other commutators vanishing. The evolution thus corresponds to a symplectic transformation generated by the underlying $SU(1,1)$ structure.



The dynamical origin of the amplified and damped branches of the doubled Bateman system can be understood from the underlying group-theoretic structure of the Hamiltonian. Defining the operators
\begin{equation}
K_+
=
a^\dagger b^\dagger,
\quad
K_-
=
ab,
\quad
K_0
=
\frac12
(a^\dagger a+b^\dagger b+1),
\end{equation}
one obtains the generators of the $SU(1,1)$ algebra, $[K_+,K_-] = -2 K_0$, $[K_0, K_{\pm}] = \pm K_{\pm}$. The Hamiltonian may then be expressed as a linear combination of these generators, and the evolution induces trajectories through the manifold of two-mode squeezed states
\begin{equation}
|0(\zeta)\rangle
=
e^{\zeta K_+-\zeta^*K_-}
|0,0\rangle,
\end{equation}
where $\zeta$ is a complex squeezing parameter
\cite{Celeghini:1991yv,RevModPhys.62.867}.

Notice that the term
\begin{equation}
H_0
\equiv
\hbar\Omega_0
(a^\dagger a-b^\dagger b)
\end{equation}
in $H$ [Eq.~(\ref{H_ab})] is proportional to the $SU(1,1)$ Casimir operator and is therefore a constant of motion,
\begin{equation}
[H_0,H_I]
=
0.
\end{equation}
Its value remains conserved throughout the evolution and, once chosen to be positive definite at the initial time, the Hamiltonian stays bounded from below.

%

%
The  evolution generated by the Bateman Hamiltonian drives the system through a family of time-dependent two-mode $SU(1,1)$ squeezed coherent states \cite{Celeghini:1991yv},
\begin{equation} \label{Ot}
|0,0\rangle
\rightarrow
|0(t) \rangle
=
e^{iHt/\hbar}|0,0\rangle
=
e^{iH_I t/\hbar}|0,0\rangle,
\end{equation}
where the notation is $|0(t) \rangle \equiv |0,0;(t)\rangle$, and we have used 
$H_0|0,0\rangle = 0$.

The evolved vacuum state $|0(t)\rangle$ is a  normalized state $\langle0(t)|0(t)\rangle = 1, ~\forall t$,  annihilated by the time-dependent operators, $e^{iH t/\hbar} a(0) e^{-iH t/\hbar}$ and $e^{iH t/\hbar} b(0) e^{-iH t/\hbar}$ (cf. Eqs. (\ref{bogo}) and (\ref{g})),
\begin{equation}
a(t)|0(t)\rangle
=
0,
\qquad
b(t)|0(t)\rangle
=
0,
\qquad
\forall t.
\label{f2}
\end{equation}
Also note that $a(t) |0,0\rangle \neq 0$ and $b(t) |0,0\rangle \neq 0$.

\section{Reduced dynamics, loss of scaling periodicity, and non-Markovian memory}

The underlying \(SU(1,1)\) structure becomes especially transparent upon introducing the collective operators
\begin{equation}
A_\pm
=
\frac{1}{\sqrt2}
\left(
a\pm b^\dagger
\right).
\label{Apm}
\end{equation}
These operators identify the amplified and damped collective directions of the Bogoliubov dynamics generated by the Bateman Hamiltonian.

Using the Heisenberg evolution [see Eqs.~(\ref{bogo}) and (\ref{g})],
\begin{align}
a(t)
&=
e^{-i\Omega_0 t}
\Big[
a\cosh(\Gamma t)
+
b^\dagger\sinh(\Gamma t)
\Big],
\nonumber\\
b(t)
&=
e^{i\Omega_0 t}
\Big[
b\cosh(\Gamma t)
+
a^\dagger\sinh(\Gamma t)
\Big],
\end{align}
one immediately obtains
\begin{equation}
A_\pm(t)
=
e^{(-i\Omega_0\pm\Gamma)t}
A_\pm(0).
\label{s}
\end{equation}
The dynamics therefore decomposes naturally into amplified and damped collective sectors characterized respectively by the exponents
\(
-i\Omega_0+\Gamma
\)
and
\(
-i\Omega_0-\Gamma.
\)

Introducing the corresponding scaling observables
\begin{equation}
Q_\pm
=
A_\pm^\dagger A_\pm,
\end{equation}
their evolution becomes
\begin{equation}
Q_\pm(t)
=
e^{\pm2\Gamma t}
Q_\pm(0).
\end{equation}
The exponential growth and decay of \(Q_\pm\) encode the self-similar scaling structure of the doubled Bateman dynamics, with \(Q_+\) and \(Q_-\) characterizing the amplified and damped collective sectors, respectively. Defining the characteristic time scale $T
=
\frac{2\pi}{\Omega_0}$, one finds
\begin{equation}
Q_\pm(t+nT)
=
e^{\pm2\Gamma nT}
Q_\pm(t),
\qquad
n\in\mathbb Z.
\label{scalinglaw}
\end{equation}
Equation~(\ref{scalinglaw}) shows that the dynamics is not periodic in the ordinary sense. Instead, after each discrete time translation $T$, the collective observables reproduce themselves up to a multiplicative scaling factor, thereby realizing a form of scaled periodicity. The full doubled Bateman system therefore exhibits a 
logarithmic-spiral self-similar structure characterized by a discrete scaling covariance.

To describe one of the 
oscillator sector, e.g., the $a$-oscillator, we introduce the reduced density matrix
\begin{equation}
\rho_a(t)
=
{\rm Tr}_b
\Big[
\rho^{\rm tot}(t)
\Big].
\end{equation}
Expectation values of subsystem observables are then given by
\begin{equation}
\langle O_a(t)\rangle
=
{\rm Tr}_a
\Big[
\rho_a(t)\,O_a
\Big].
\end{equation}
We note that here and in the following expectation values are computed in the $SU(1,1)$ time dependent state $|0(t)\rangle$.
The observables $Q_\pm$ responsible for the scaling covariance are 
%
\begin{equation}
Q_\pm
=
\frac12
\Big[
N_a+N_b+1
\pm
\left(
a^\dagger b^\dagger
+
ab
\right)
\Big],
\label{Qpmdecomp}
\end{equation}
where $N_a = a^\dagger a$, $N_b = b^\dagger b$, 
showing 
that $Q_\pm$ involve both oscillator sectors through the $N_a$ and $N_b$, and the cross-sector correlations $ab$ and $a^\dagger b^\dagger$. Consequently, $Q_\pm$ cannot be reconstructed solely from the reduced density matrix $\rho_a$.

The loss of scaling covariance at the subsystem level therefore does not occur because the symmetry of the full system is destroyed.  
As a result, for a generic observable $O_a$ of the reduced subsystem, we have
\begin{equation} \label{Q}
\langle O_a(t+nT)\rangle
\neq
e^{\lambda nT}
\langle O_a(t)\rangle,
\qquad
n\in\mathbb Z.
\end{equation}

The information associated with the scaling structure is nevertheless not lost. Since the full evolution remains unitary, the correlations 
in for Eq.~(\ref{Qpmdecomp}) remain encoded in the full density matrix $\rho^{\rm tot}(t)$.

The reduced dynamics derived in Appendix~A is governed by (cf. Eqs. (\ref{eom_bateman}) -  (\ref{Kb}))
\begin{equation}
\frac{d}{dt}\rho_a(t)
=
\mathcal L_u\rho_a(t)
+
\mathcal K[\rho^{\rm tot}(t)],
\end{equation}
where
\begin{equation}
\mathcal K[\rho^{\rm tot}(t)]
=
-\frac{1}{\hbar^2}
\int_0^t ds\,
{\rm Tr}_b
\Big[
H_{\rm int}^{I}(t),
[H_{\rm int}^{I}(s),
\rho^{\rm tot}(s)]
\Big].
\end{equation}
Consequently, the time evolution of the reduced density matrix ($\rho_a(t)$) depends explicitly on earlier states through the memory kernel governing the reduced dynamics. The subsystem evolution therefore retains information about past correlations generated by the interaction between the oscillator sectors.

Unlike conventional Markovian dissipation, where information is irreversibly lost into a large reservoir, the doubled Bateman system preserves the correlations responsible for the amplified and damped branches of the full dynamics.  
These collective structures 
survive through the history dependence of the reduced evolution. The resulting subsystem dynamics is therefore intrinsically non-Markovian. More importantly, the same cross-sector correlations that generate the amplified and damped collective structures of the full doubled system continue to influence the observable sector through the memory kernel governing the reduced evolution.

The detailed derivation of the reduced evolution equation, the associated memory kernel, and the corresponding subsystem observables is presented in Appendix~A.

A crucial observation is that the memory kernel depends on the complete history of the full density matrix through $\rho^{\rm tot}(s)$. The future evolution of the subsystem is therefore influenced by correlations generated at earlier times.

Furthermore, Appendix~A shows that the same cross-sector operators
$ab$ and $a^\dagger b^\dagger$,
which contribute to the collective observables $Q_\pm$, also enter explicitly into the memory kernel. Thus, the operator content responsible for the self-similar scaling structure of the full doubled system survives within the reduced dynamics even though the corresponding scaling observables are not directly accessible.

The physical picture is therefore the following. In the full doubled system, the amplified and damped sectors combine to form collective observables exhibiting exact scaling covariance. After tracing over one sector, these observables become inaccessible and the scaling law is no longer visible at the level of subsystem expectation values. Nevertheless, the correlations responsible for the scaling structure remain present in the full state and re-enter the observable dynamics through the memory kernel.

In this sense, the partial trace converts an explicitly observable scaling covariance into information stored in cross-sector correlations. The scaling law itself is not recovered at the subsystem level. Instead, its dynamical influence survives indirectly through non-Markovian memory. The memory kernel therefore acts as the mechanism through which information associated with the amplified and damped self-similar branches continues to affect the future evolution of the observable sector.

This observation is central for understanding the relation to the no-go theorem of Watanabe and Oshikawa \cite{Watanabe:2014hea}. The theorem assumes a closed equilibrium system and concerns the behavior of directly observable local operators. In contrast, the subsystem considered here is obtained through a partial trace over an interacting amplified sector and is governed by a non-Markovian evolution equation. Moreover, the relevant symmetry of the full doubled dynamics is a scaling covariance of collective observables rather than an ordinary temporal periodicity of local observables.

In the geometric and dynamical picture developed above, 
although the emergence of scaled periodicity and non-Markovian memory is already visible at the level of the doubled oscillator system, the underlying $SU(1,1)$ structure acquires additional significance in the QFT infinite-volume limit.

In this limit, the time evolution generated by the squeezing interaction develops moving through a continuous family of unitarily inequivalent representations of the canonical commutation relations. For the vacuum states $|0(t)\rangle$, $\forall t$, 
%
%
%
in the infinite-volume limit one finds
\begin{equation}
\langle0(t')|0(t)\rangle
\rightarrow
0,
\qquad
V\rightarrow\infty,
\qquad
t'\neq t.
\label{uir}
\end{equation}
Equation~(\ref{uir}) expresses the well-known QFT feature that the vacuum states $|0(t)\rangle$ at 
different times become orthogonal in the infinite-volume limit and therefore belong to unitarily inequivalent representations of the canonical commutation relations. 
The doubled Bateman dynamic time evolution is thus shown to be associated with a continuous sequence of time-dependent vacuum states, reflecting its intrinsically nonequilibrium character \cite{Celeghini:1991yv}. Similar connections between time evolution, Bogoliubov transformations, and families of time-dependent vacuum states have recently been discussed in the context of Hamiltonian time-crystal constructions \cite{sytq-kjbr}.

As already observed, the generalized $SU(1,1)$ squeezed coherent states states $|0(t)\rangle$ \cite{Perelomov} 
exhibit entanglement between the two oscillator modes \cite{Gerry}, and provide the proper framework of nonequilibrium finite-temperature quantum field theory \cite{Celeghini:1991yv,Umezawa1982}.

The geometric representation of the nonequilibrium dynamics follows directly from the classical independent modes of the Bateman system, without introducing any additional dynamical assumptions. As implied by Eqs.~\eqref{bogo} and \eqref{g}, the general solution can be decomposed into the elementary modes
\begin{equation}
z_1(t)
=
A_1
e^{-\Gamma t}
e^{-i\Omega_{0} t},
\qquad
z_2(t)
=
A_2
e^{+\Gamma t}
e^{+i\Omega_{0} t},
\end{equation}
forming a complete basis for the classical dynamics. The physical variables $Y_1(t)$ and $Y_2(t)$ may therefore be reconstructed as linear combinations of these modes and their complex conjugates, reflecting the completeness of the exponential basis $e^{(\pm\Gamma\pm i\Omega_{0})t}.$
Geometrically, each mode combines rotational motion generated by the oscillatory phase factor with exponential contraction or expansion governed by the damping parameter $\Gamma$. The corresponding trajectories in the complex plane take the form of logarithmic spirals, corresponding respectively to the damped and amplified sectors of the theory.

The classical logarithmic-spiral trajectories admit a natural quantum counterpart through the collective modes defined in Eq.~(\ref{s}), which encode the amplified and damped branches of the doubled quantum dynamics. 

The discrete scaled periodicity described by Eq.~(\ref{scalinglaw}) may then be understood as the quantum realization of the logarithmic-spiral self-similar structure already inherent in the classical Bateman system.
The non-Markovian evolution of the reduced subsystem may be traced to the same cross-sector correlations that underlie the amplified and damped logarithmic-spiral branches of the full doubled Bateman dynamics.

\section{Fractal self-similarity and time-crystals}

In the previous sections, we have shown that the quantum dynamics of the Bateman system is governed by Bogoliubov transformations and $SU(1,1)$ squeezed coherent-state structures. In this section, we provide a {\it geometric representation} of these results, showing that the same dynamics can be understood in terms of fractal self-similarity and logarithmic spiral trajectories \cite{Vitiello2012}. This connection clarifies the origin of the discrete lattice-like temporal structures observed at the quantum level, providing the realization of time-crystal--like behavior.

The most important property of fractals is their self-similarity \cite{Peitgen}. A paradigmatic example is the logarithmic spiral, defined in polar coordinates $(r,\phi)$ by
\begin{equation}
r = r_0 \, e^{d \, \phi},
\end{equation}
with $r_0 > 0$ and $d$ a real constant \cite{Peitgen,Andronov,Vitiello2012}. The self-similarity relation
\begin{equation}
d \, \phi = \ln \frac{r}{r_0}
\end{equation}
shows that a shift $\phi \rightarrow \phi' =  \phi + \Delta$ rescales $r$ by a constant factor $e^{d\Delta}$, $r \rightarrow r' = r\, e^{d\Delta} $ leaving the shape invariant.

The spiral can be represented in the complex plane as
\begin{equation}
z = x + i y = r_0 \, e^{d \, \phi} \, e^{i \, \phi},
\end{equation}
with
\begin{eqnarray}
x &=& r_0 e^{d\phi}\cos\phi, \\
y &=& r_0 e^{d\phi}\sin\phi.
\end{eqnarray}

Note that the spiral is fully specified only upon fixing the sign of $d\phi$. The completeness of the basis $\{e^{-d\phi}, e^{+d\phi}\}$ therefore requires that both factors $q = e^{\pm d\phi}$ be considered, corresponding to the two possible chiralities: the `direct' (left-handed, $q>1$) and the `indirect' (right-handed, $q<1$) spirals (see Fig.~2). Interestingly, both contracting and expanding spirals are observed in nature, for instance in phyllotaxis.

We are thus naturally led to consider
\begin{equation}
z_1 = r_0 \, e^{- d \phi} \, e^{- i \phi}, 
\qquad 
z_2 = r_0 \, e^{+ d \phi} \, e^{+ i \phi}.
\end{equation}

\begin{figure}[t]
\centering
\includegraphics[width=0.85\columnwidth]{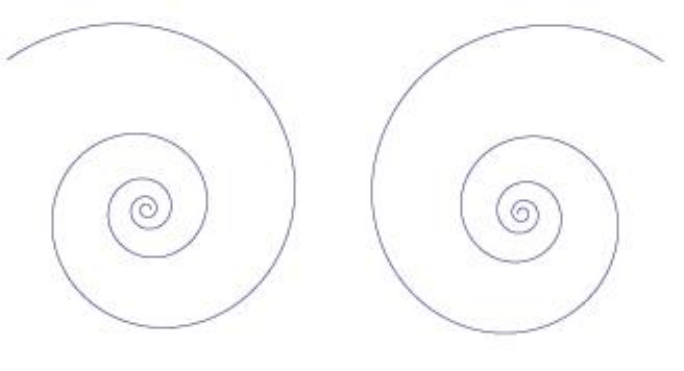}
\caption{
Trajectory in the complex plane illustrating the logarithmic spiral structure arising from the combined oscillatory and scaling dynamics.
}
\label{fig2}
\end{figure}

To connect this geometric construction with dynamics, we promote $\phi$ to a time-dependent variable:
\begin{equation}
\phi(t) = \frac{\Gamma}{d}\, t,
\end{equation}
where $\Gamma$ is defined in Eq.~(\ref{Gamma}).

This choice is a parametrization matching the spiral geometry with the dynamical evolution. 
The relation $\Omega_{0} = \Gamma/d$ follows from this mapping and expresses
the ratio between damping and oscillation in the spiral geometry.
The parameter $d$ plays a central role.
It acts as a scaling exponent controlling the rate
of self-similar growth or decay of the logarithmic spiral. 
It encodes how the interplay between dissipative and oscillatory dynamics gives rise to the observed self-similar (fractal-like) scaling structure. Substituting into $z_1$ and $z_2$, we get
\begin{equation}
z_1(t) = r_0 \, e^{-\Gamma t} e^{-i\Omega_{0} t}, 
\qquad 
z_2(t) = r_0 \, e^{+\Gamma t} e^{+i\Omega_{0} t}.
\end{equation}

These expressions solve the Bateman equations
\begin{eqnarray}
\ddot{z}_1 + \gamma \dot{z}_1 + \Omega^2 z_1 &=& 0, \\
\ddot{z}_2 - \gamma \dot{z}_2 + \Omega^2 z_2 &=& 0,
\end{eqnarray}
with $\Omega_0^2=\Omega^2-\Gamma^2$ and $\gamma=2\Gamma$.
Separately, $z_1$ and $z_2$ describe open systems. Taken together, however, they form a closed system, reflecting the same doubling of degrees of freedom encountered in the quantum description.

From $\phi(T)=2\pi$, one finds
\begin{equation}
T = \frac{2\pi}{\Omega_{0}}.
\end{equation}

However, the evolution is not strictly periodic. Instead,
\begin{equation}
z_i(t+T) = z_i(t) e^{\mp 2\pi d}.
\label{re}
\end{equation}

\begin{figure}[t]
\centering
\begin{tikzpicture}
\begin{axis}[
    width=\columnwidth,
    height=0.6\columnwidth,
    xlabel={$n$},
    ylabel={$r(nT)$},
    ymode=log,
    grid=both,
    title={Discrete-time scaling},
    tick label style={font=\small},
    label style={font=\small},
    colormap/viridis,
    colorbar,
    point meta min=0,
    point meta max=9,
    colorbar style={title={$n$}}
]

\addplot[
    scatter,
    only marks,
    scatter src=explicit,
    mark=*,
    mark size=2.5pt,
    thick
]
coordinates {
(0,1.000) [0]
(1,0.286) [1]
(2,0.082) [2]
(3,0.023) [3]
(4,0.0067) [4]
(5,0.0019) [5]
(6,0.00055) [6]
(7,0.00016) [7]
(8,0.000045) [8]
(9,0.000013) [9]
};

\addplot[
    thick,
    black,
    domain=0:9,
    samples=100
]
{exp(-1.25*x)};

\addlegendentry{$r(nT)=e^{-\Gamma nT}$}

\end{axis}
\end{tikzpicture}
\caption{
Discrete-time sampling of $r(t)$ at $t=nT$. The color gradient
indicates increasing $n$. The linear behavior in logarithmic scale
reveals the underlying self-similar (fractal-like) scaling and the
associated time-lattice structure.
}
\label{fig:scaling}
\end{figure}
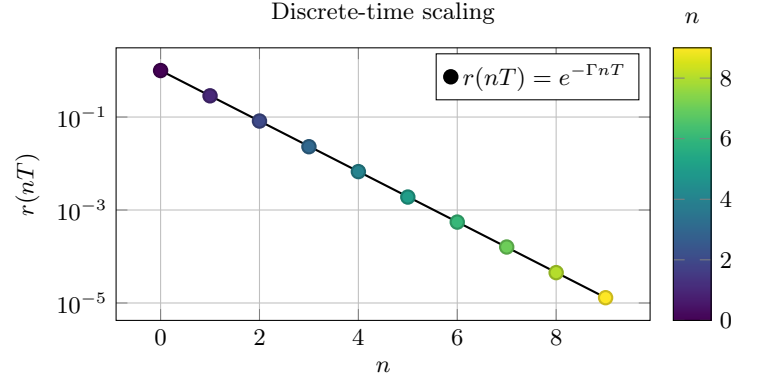

The discrete-time structure of the system becomes manifest when
sampling the dynamics at intervals $t=nT$. As shown in
Fig.~\ref{fig:scaling}, the quantity $r(nT)$ follows an exponential
scaling law, which appears as a linear behavior in logarithmic scale,
revealing the underlying self-similar (fractal-like) structure.

Thus, time evolution generates scaled copies of the trajectory rather than exact repetitions. This defines a discrete scaling structure in time, characterized by the scaling interval  (``the time-lattice'') $\ell_d=T$ .

For $t=nT$,
\begin{equation}
z_1(nT)=r_0 e^{-2\pi d n},
\qquad
z_2(nT)=r_0 e^{+2\pi d n},
\end{equation}
or equivalently,
\begin{equation} \label{scrat}
\frac{r_i(t_n+T)}{r_i(t_n)}
=
e^{\mp 2\pi d},
\end{equation}
which may be rewritten in logarithmic form as
\begin{equation}
\ln r_i\!\big((n+1)\ell_d\big)
=
\ln r_i(n\ell_d)
\mp 2\pi d .
\end{equation}
The scaling exponent is given by
\begin{equation}
d=
\frac{\Gamma}
{\sqrt{\Omega^2-\Gamma^2}}
=
\frac{\Omega s}{\sqrt{4m^{2}-\Omega^{2}s^{2}}},
\end{equation}
where the second equality follows from Eq.~\eqref{Gamma}. This expression explicitly identifies the fractional spin parameter $s$ as the physical origin of the scaling exponent $d$ and, consequently, of the discrete scale covariance exhibited by the dynamics. Since the scaling ratio introduced above is completely determined by $d$, the strength of the discrete scaling covariance is directly controlled by the underlying fractional spin. The discrete temporal scaling therefore constitutes a direct dynamical manifestation of the spin-induced deformation and disappears continuously in the spinless limit $s\to0$, where $d\to0$ and the scaling ratio  Eq. (\ref{scrat}) approaches unity.

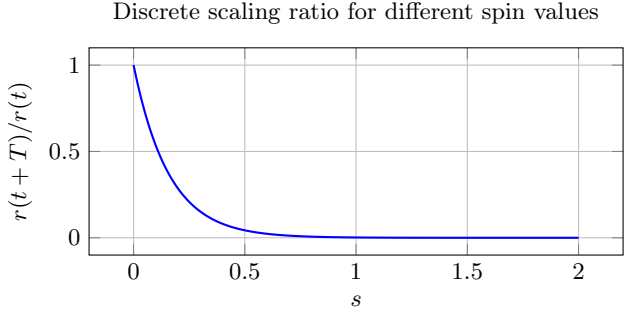
\begin{figure}[t]
\centering
\begin{tikzpicture}
\begin{axis}[
    width=\columnwidth,
    height=0.5\columnwidth,
    xlabel={$s$},
    ylabel={$r(t+T)/r(t)$},
    grid=both,
    title={Discrete scaling ratio for different spin values},
    tick label style={font=\small},
    label style={font=\small}
]

\addplot[
    thick,
    blue,
    domain=0:2,
    samples=200
]
{exp(-2*pi*x)}; 

\end{axis}
\end{tikzpicture}
\caption{
The ratio $r(t+T)/r(t)=e^{-2\pi d}$ as a function of the parameter $s$
(with $d \propto s$). The dependence illustrates how the discrete
scaling structure is controlled by the underlying dynamical parameter.
}
\label{fig:ratio}
\end{figure}

The existence of this constant spin-controlled scaling factor demonstrates that the dynamics is invariant under discrete temporal rescalings. This feature is further illustrated in Fig.~\ref{fig:ratio}, where the ratio $r(t+T)/r(t)$ remains constant throughout the evolution, providing direct evidence for discrete scale covariance in time \cite{Giergiel:2018slx}.

In fact, this property bears a formal resemblance to the Bloch
relation in spatial crystals,
\begin{equation}
\psi(x+\epsilon)=e^{ik\epsilon}\psi(x),
\end{equation}
where a discrete translation changes the state only by a
multiplicative factor. Similar space-time analogies have
been extensively discussed in the context of crystalline
structures emerging in the time domain
\cite{CeleghDeMart:1995,PhysRevA.98.013613}. In the present case, however, the relevant transformation is not a spatial translation but a discrete temporal rescaling, as follows from Eq.~(\ref{re}). Consequently, evolution over one scaling period generates a self-similar copy of the trajectory differing only by an overall scale factor.

From this perspective, the dynamics displays a temporal analogue of the Bloch property, with the complex Bloch phase $(e^{ik\epsilon})$ replaced by the real scaling factor $(e^{\mp 2\pi d})$. The underlying symmetry is therefore fundamentally different from ordinary temporal periodicity, where the trajectory reproduces itself exactly after a fixed period. Instead, the system exhibits a discrete scaling covariance: each interval of duration (T) recreates the same dynamical pattern up to a constant rescaling. This self-similar structure reflects the coexistence of the exponentially amplified and damped branches that are intrinsic to the Bateman system.



Thus, periodic behavior appears at the level of observables, even though the underlying dynamics is characterized by discrete scaling rather than strict periodicity. This feature provides a time-crystal--like structure in the sense of persistent temporal ordering through definite lattice--like time-translations. {\it The growth of self-similar fractal structures represents therefore the observable manifestation of crystal--like time ordering}. 

\medskip

For completeness, we extend this analysis to other fractals, such as the Koch curve \cite{Peitgen}. Let $u_{n,q}(\alpha)$ denote the $n$-th step of the Koch construction, with $\alpha = 4$ and scaling parameter $q = 1/3^d$. The iterative relation
\begin{equation}
u_{n,q}(\alpha) = (q\alpha)^n
\end{equation}
leads to the fractal dimension $d = \ln 4 / \ln 3$. Generalizing the $\alpha$ and $q$, and  introducing the parametrization \cite{Vitiello2012} $q = e^{-d\phi}$, one obtains
\begin{equation}
u = \alpha e^{d\phi},
\end{equation}
which has the same functional form as the logarithmic spiral. The corresponding self-similarity relation is
\begin{equation}
d\phi = \ln \frac{u}{\alpha}.
\end{equation}

This shows that the Koch fractal can be described within the same framework as the logarithmic spiral (see Fig.~\ref{fig2}). In particular, its scaling properties can be mapped onto a pair $(z_1, z_2)$ through the doubling of degrees of freedom, leading once again to a discrete time structure with $\ell_d \propto d/\Gamma$.

Within this picture, the parameter $q = e^{-d\phi}$ plays the role of a deformation parameter, establishing a direct link with $q$-deformed coherent states \cite{Vitiello2012}. The fractal dimension $d$ encodes the scaling properties and is directly related to the squeezing parameter appearing in the quantum description.

At the quantum level, the doubling of the degrees of freedom,
${\cal A} \rightarrow {\cal A}_1 \otimes {\cal A}_2$,
is associated with the noncommutative Hopf algebra coproduct
\begin{equation}
{\cal A} \rightarrow {\cal A} \otimes \mathbf{q} + \mathbf{q^{-1}} \otimes {\cal A},
\end{equation}
from which the Bogoliubov transformations are derived \cite{CeleghDeMart:1995,Celeghini:1998a}. 
The deformation parameter $q$ is thus related to the squeezed coherent condensate characterizing the state $|0(d,t)\rangle$ \cite{Vitiello2012}. A change in $q$, induced by a variation of the fractal dimension $d$ (i.e., the slope in the log-log representation), corresponds to a transition between unitarily inequivalent representations,
\begin{equation} \label{qdef}
|0(d,t)\rangle \;\longrightarrow\; |0(d',t)\rangle, \qquad d \neq d'.
\end{equation}

\begin{figure}[t]
\centering
\includegraphics[width=0.47\linewidth]{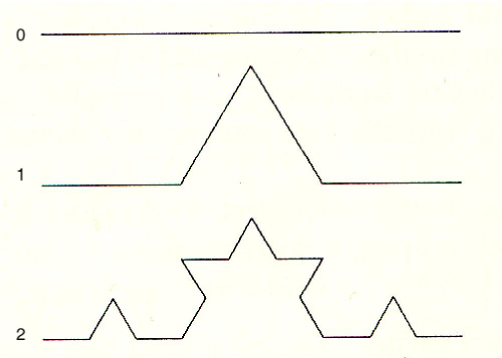}
\hfill
\includegraphics[width=0.47\linewidth]{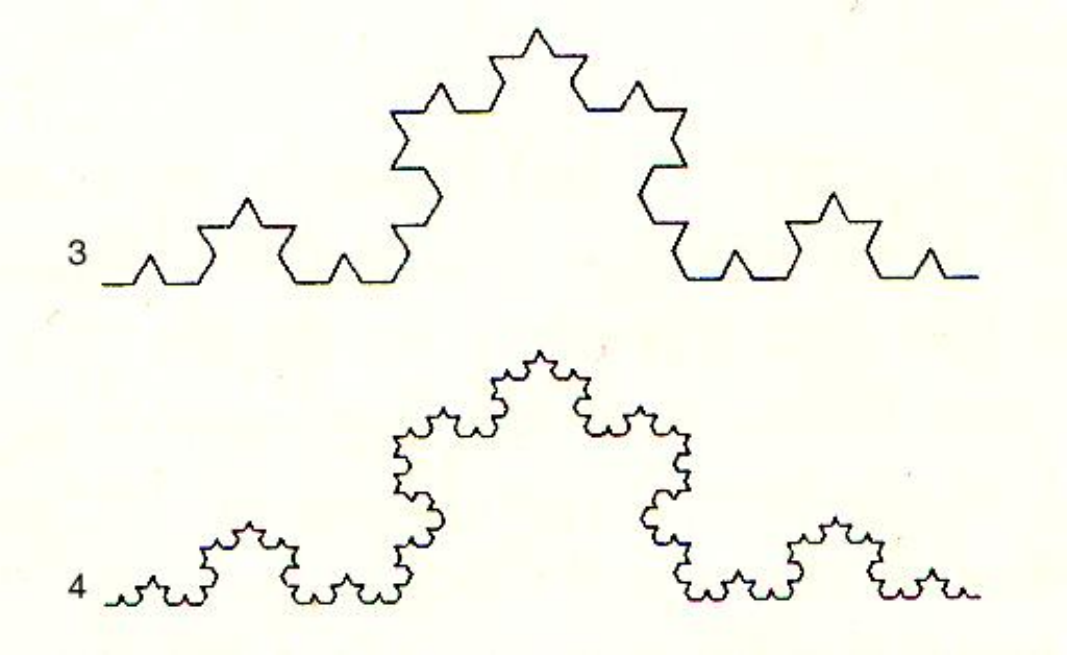}
\caption{
\small
Successive stages in the construction of the Koch
curve, illustrating the emergence of self-similar
fractal structure.
}
\label{fig5}
\end{figure}

This provides a physical interpretation of the interplay between dissipation (at the origin of the $q$-deformation of the Hopf algebra), noncommutative geometry, and the nontrivial topology of trajectories in phase space \cite{Celeghini:1998a,Iorio:1994jk,Sivasubramanian:2003xy}. For further details, we refer to the cited literature.

The logical structure of our results and the chain of connections can be schematically represented as in Fig.~\ref{df}.

\section{Concluding remarks, interpretation and outlook}

In this work, we investigated the quantum Bateman oscillator within a phase-space noncommutative framework governed by Dirac brackets. After quantization, the system is described by a time-independent Hermitian Hamiltonian whose dynamics is naturally organized in terms of amplified and damped collective modes associated with an underlying $\mathrm{SU}(1,1)$ structure. We showed that the full doubled system possesses an exact self-similar scaling structure characterized by a discrete scaling periodicity, while the corresponding reduced subsystem exhibits intrinsically non-Markovian dynamics governed by a history-dependent memory kernel.

\begin{figure}[t]
\centering
\begin{tikzpicture}[
node distance=0.9cm,
every node/.style={
draw,
rounded corners,
align=center,
minimum width=3.5cm,
minimum height=0.8cm,
font=\small
},
arrow/.style={->, thick, >=Stealth}
]

\node (n1) {Nonrelativistic spin ($s\neq 0$)};
\node (n2) [below=of n1] {Spin-induced deformation ($\theta=s/m^2$)};
\node (n3) [below=of n2] {Quantum deformed phase space};
\node (n4) [below=of n3] {Doubled Bateman dynamics};
\node (n5) [below=of n4] {Amplified / damped modes};
\node (n6) [below=of n5] {Logarithmic-spiral\\self-similarity};
\node (n7) [below=of n6] {Self-similar scaling\\periodicity};
\node (n8) [below=of n7] {Reduced dynamics};
\node (n9) [below=of n8] {Non-Markovian memory};

\draw[arrow] (n1) -- (n2);
\draw[arrow] (n2) -- (n3);
\draw[arrow] (n3) -- (n4);
\draw[arrow] (n4) -- (n5);
\draw[arrow] (n5) -- (n6);
\draw[arrow] (n6) -- (n7);
\draw[arrow] (n7) -- (n8);
\draw[arrow] (n8) -- (n9);

\end{tikzpicture}
\caption{
Schematic flow of the underlying mechanism. Nonrelativistic spin induces a noncommutative phase-space structure, which leads to the doubled Bateman dynamics. The amplified and damped collective modes generate logarithmic-spiral self-similarity and an exact self-similar scaling periodicity in the full doubled system. After tracing over one oscillator sector, the explicit scaling structure becomes inaccessible, but its influence survives through the memory kernel governing the resulting non-Markovian reduced dynamics.
}
\label{df}
\end{figure}
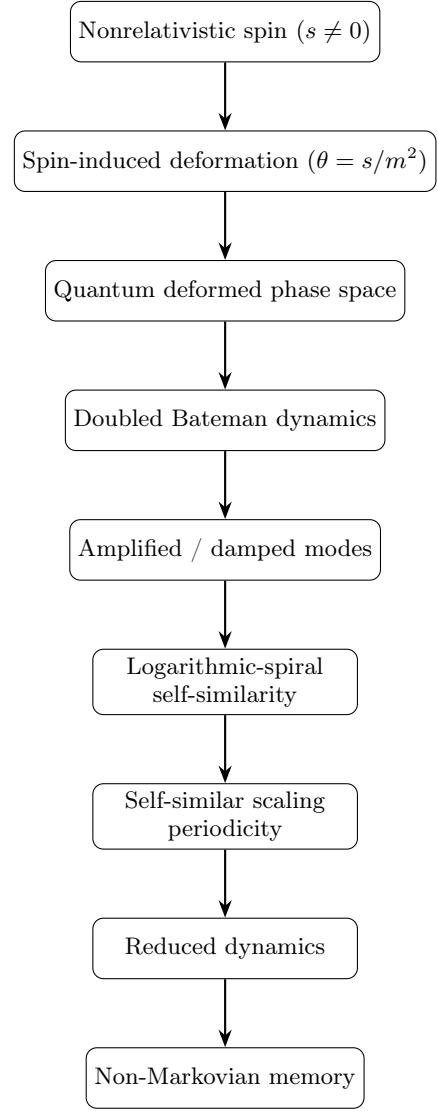

We further demonstrated that the Bateman phase-space dynamics admits a natural geometric realization in terms of self-similar fractal structures, including logarithmic spirals and Koch-type scaling curves (see Fig.~\ref{fig5}), thereby revealing a close connection between geometric scaling properties and the amplified and damped branches of the underlying dynamics. A central aspect of the present framework concerns the origin of the noncommutative structure. The parameter $\theta$ is introduced dynamically rather than as a fundamental deformation. In the first-order formulation, the dissipative structure emerges from a noncanonical symplectic geometry which, upon implementation of the Dirac bracket formalism, naturally leads to noncommuting coordinates. The resulting noncommutativity can therefore be traced back to underlying nonrelativistic spin degrees of freedom within a definite physical setting (see Fig.~\ref{df}).

The Bateman oscillator provides a paradigmatic framework for dissipative dynamics, amplification/damping, doubled degrees of freedom, and nonequilibrium evolution, with established connections to noncommutative geometry, Bogoliubov transformations, and nonequilibrium quantum field theory. Unlike conventional scenarios based on spontaneous symmetry breaking or external Floquet driving, the self-similar scaling periodicity identified here emerges directly from the intrinsic dynamics of the doubled Bateman system itself.

\begin{table}[t]
\centering
\caption{
Key features of the present framework.
}
\begin{tabular}{|c|c|}
\hline
Hamiltonian
&
Time independent
\\
\hline

Underlying structure
&
$SU(1,1)$ doubled dynamics
\\
\hline

Collective modes
&
Amplified and damped branches
\\
\hline

Scaling law
&
$Q_\pm(t+nT)=e^{\pm2\Gamma nT}Q_\pm(t)$
\\
\hline

Subsystem dynamics
&
Non-Markovian
\\
\hline

Memory effects
&
History dependent
\\
\hline

Global evolution
&
Unitary
\\
\hline

Observable scaling periodicity
&
Lost after partial trace
\\
\hline

Hidden structure
&
Preserved through correlations
\\
\hline

Geometric interpretation
&
Logarithmic-spiral self-similarity
\\
\hline

No-go theorem
&
Outside equilibrium assumptions
\\
\hline

\end{tabular}
\label{tab:comparison}
\end{table}

After mapping the system to canonical variables and proceeding to quantization, the resulting Hamiltonian describes two coupled oscillator sectors governed by an underlying $\mathrm{SU}(1,1)$ dynamical structure. The corresponding amplified and damped collective modes evolve according to $A_{\pm}(t)=e^{(-i\Omega_0\pm\Gamma)t}A_{\pm}(0)$ and give rise to collective observables $Q_\pm$, which characterize the strength of the amplified and damped branches of the doubled dynamics. These observables satisfy the exact scaling relations $Q_{\pm}(t+nT)=e^{\pm2\Gamma nT}Q_{\pm}(t)$, with $n\in\mathbb Z$, demonstrating that the full doubled system possesses a discrete self-similar scaling covariance. Time evolution therefore generates a sequence of self-similar scaled copies rather than exact periodic repetitions, revealing a temporal scaling structure that emerges entirely from the intrinsic dynamics of the doubled Bateman system, without external driving or Floquet engineering.

An important aspect of the present mechanism is that the observable oscillator sector admits a natural reduced description obtained by tracing over one oscillator degree of freedom. The resulting reduced dynamics is intrinsically non-Markovian because the evolution equation contains a memory kernel that depends explicitly on the interaction history.

Unlike conventional Markovian dissipation, where information is irreversibly lost into a macroscopic reservoir, the doubled Bateman framework preserves the correlations responsible for the amplified and damped branches of the full dynamics. Although the collective observables generating the scaling structure become inaccessible after the partial trace, the corresponding correlations remain encoded in the full density matrix and continue to influence the observable subsystem through the memory kernel.

Consequently, the reduced dynamics retains an indirect imprint of the self-similar scaling structure present in the full doubled system. In this sense, non-Markovian memory plays a central role in transmitting the influence of the amplified and damped sectors to the observable subsystem.

In terms of complex modes, the classical dynamics takes the form $z(t)\sim e^{-\Gamma t}e^{-i\Omega t}$, combining exponential scaling with rotational motion. This structure naturally gives rise to logarithmic-spiral trajectories whose self-similar properties provide a geometric manifestation of the amplified and damped dynamics. The corresponding quantum description emerges through the collective modes $A_{\pm}$, whose scaling behavior may be viewed as the quantum counterpart of the logarithmic-spiral structure already present in the classical Bateman system. It is interesting to note that connections between temporal ordering and fractal self-similar structures have also recently been identified in higher-order Floquet time crystals \cite{PhysRevB.108.L140102}. In the present case, however, the self-similarity emerges directly from the amplified and damped dynamics of the Bateman system and is encoded in the dynamical trajectories themselves rather than in the phase structure of a periodically driven system.

In contrast to scenarios based on external driving, the behavior identified here arises entirely from the internal structure of the system. The relevant dynamical feature is not an ordinary temporal periodicity but an exact self-similar scaling structure of collective observables in the full doubled dynamics. Time evolution therefore generates a sequence of self-similar scaled copies rather than exact periodic repetitions, revealing a form of temporal ordering associated with discrete scaling covariance.

It is important to emphasize that the present mechanism does not correspond to the spontaneous breaking of continuous time-translation symmetry within a single representation of the canonical commutation relations in the sense addressed by the no-go theorem of Watanabe and Oshikawa~\cite{Watanabe:2014hea}. Rather, the self-similar scaling structure originates from the amplified and damped collective modes of the doubled Bateman system and survives in the reduced description only through the history-dependent memory kernel governing the non-Markovian evolution of the observable subsystem.

The non-Markovian reduced dynamics discussed above originates from a full unitary evolution taking place in the enlarged doubled Hilbert space. The nonequilibrium character of this underlying dynamics is reflected in the structure of the corresponding vacuum states. In the infinite-volume limit, the time evolution generated by the $\mathrm{SU}(1,1)$ interaction connects a family of unitarily inequivalent representations of the canonical commutation relations. For clarity, some of the central conceptual features of
the present framework are summarized in
Table~\ref{tab:comparison}.

The present framework also suggests experimentally accessible signatures.  
As mentioned, the influence of self-similar scaling periodicity survives through the memory kernel governing the reduced evolution. Consequently, the most direct experimental signatures of the underlying amplified and damped structure are expected to appear  
through measurable non-Markovian effects such as information backflow, temporal correlations, and memory-induced deviations from Markovian relaxation.

Platforms supporting engineered gain-loss dynamics, including coupled photonic resonators, optomechanical systems, superconducting circuits, and trapped-ion quantum simulators, provide promising settings in which to investigate these phenomena \cite{Blatt2012,Clarke2008,Devoret2013}. In such systems, one may explore whether self-similar scaling structures manifest themselves through observable non-Markovian dynamics, thereby providing an experimental route to probe the interplay between amplified and damped collective modes, memory effects, and emergent scaling behavior. The relevance of the present results lies in providing a unified mechanism through which dissipation, noncommutative geometry, self-similar scaling structures, and non-Markovian memory can be understood within a unified setting. In particular, we have shown that dissipative subsystem dynamics can emerge consistently from a globally unitary quantum theory formulated on an enlarged doubled Hilbert space. At the same time, the noncommutative structure results naturally from the underlying dynamics rather than being imposed externally.

From a broader perspective, the present framework may be relevant to open quantum systems, quantum thermodynamics, and nonequilibrium dynamics, where dissipation is typically associated with nonunitary evolution. Our analysis demonstrates that dissipative and nonequilibrium subsystem behavior can arise consistently from an underlying globally unitary dynamics through the reduced description of coupled doubled degrees of freedom. In addition, the geometric fractal-like description provides a useful diagnostic for identifying self-similar scaling structures in dynamical evolution, with potential connections to quantum optics, coupled oscillator systems, and effective models based on noncommutative geometry.

In summary, the central result of the present work is the identification of an exact self-similar scaling structure in the doubled Bateman dynamics. Unlike ordinary temporal periodicity, time evolution generates a sequence of scaled copies of the same dynamical pattern, leading to a discrete scaling covariance of the collective observables. This scaling structure originates from the amplified and damped sectors associated with the underlying $SU(1,1)$ dynamics and emerges without external driving or Floquet engineering. Although the corresponding scaling observables become inaccessible after tracing over one oscillator sector, the correlations responsible for them survive through the memory kernel governing the reduced evolution. The resulting non-Markovian dynamics therefore retains an indirect imprint of the underlying scaling structure.


An intriguing direction for future research lies in extending the present framework to relativistic ((2+1))-dimensional systems with nontrivial spin structure, where anyonic excitations naturally arise. Since the spin parameter is responsible for the amplified and damped Bateman dynamics and the associated self-similar scaling behavior, it is natural to investigate whether analogous mechanisms persist within relativistic anyon sectors. In particular, it would be interesting to determine whether the logarithmic-spiral scaling structure survives in a Lorentz-covariant setting and how it is modified by relativistic wave-packet effects, where different momentum components experience distinct time-dilation factors. Such a study could clarify the interplay between spin-induced scaling, relativistic dephasing, and non-Markovian memory, potentially establishing a bridge between the present nonrelativistic construction and relativistic anyon wave-packet dynamics \cite{Majhi:2019etz}. We leave this interesting direction for future investigation.


\appendix

\section{Reduced dynamics of a single oscillator sector}

In this appendix, we derive the reduced non-Markovian
dynamics of one oscillator sector of the doubled
Bateman system by tracing over its coupled partner
mode. In the present discussion, we focus on the
damped oscillator sector, although an equivalent
construction may also be formulated by tracing over the
damped mode and studying the amplified sector instead.

The total Hamiltonian is
\begin{equation}
\hat H
=
\hat H_0+\hat H_{\rm int} = \hat H_a-\hat H_b -i\hbar\Gamma
(\hat a\hat b-\hat a^\dagger\hat b^\dagger),
\label{Hab} 
\end{equation}
where
\begin{equation}
\hat H_a
=
\hbar\Omega_0\,\hat a^\dagger\hat a,
\qquad
\hat H_b
=
\hbar\Omega_0\,\hat b^\dagger\hat b,
\end{equation}

The opposite signs entering $\hat H_0 =  \hat H_a-\hat H_b $ characterize the
damped and amplified sectors of the globally unitary
Bateman system. In particular, the sign structure is
not introduced by hand, but follows consistently from
the doubled Bateman dynamics and the underlying
$SU(1,1)$ algebraic structure. The quantity
$\hat H_0$ is proportional to the $SU(1,1)$ Casimir
operator and therefore defines a conserved quantity of
the dynamics (energy positiveness is preserved once positive boundary conditions are assigned at the initial time).

To study the effective subsystem evolution, we construct
a reduced description by tracing over the amplified
sector while retaining the damped oscillator as the
subsystem of interest. This tracing procedure does not
imply that the amplified sector is unphysical; rather,
it provides an effective description of the subsystem
dynamics generated by coherent exchange within the full
doubled Bateman framework.

Passing to the interaction picture with respect to
$\hat H_0$, the interaction-picture Hamiltonian becomes
\begin{align}
\hat H_{\rm int}^{I}(t)
&=
e^{\frac{i}{\hbar}\hat H_0 t}
\hat H_{\rm int}
e^{-\frac{i}{\hbar}\hat H_0 t}.
\end{align}

Using  $[\hat H_{0},\hat H_{\rm int}] = 0$, 
one obtains
\begin{align}
\hat H_{\rm int}^{I}(t)
&=
-i\hbar\Gamma
\left(
\hat a\hat b
-
\hat a^\dagger\hat b^\dagger
\right)
=
\hat H_{\rm int}.
\label{HintI}
\end{align}

Thus, the phase factors cancel because of the opposite
sign structure entering $\hat H_0$, and consequently
the interaction-picture Hamiltonian remains time
independent.

The total density matrix satisfies the von--Neumann
equation
\begin{equation}
\frac{d}{dt}\hat\rho^{\rm tot}(t)
=
-\frac{i}{\hbar}
[\hat H_{\rm int}^{I}(t),\hat\rho^{\rm tot}(t)].
\label{VN_B}
\end{equation}

Formally integrating Eq.~\eqref{VN_B} gives
\begin{equation}
\hat\rho^{\rm tot}(t)
=
\hat\rho^{\rm tot}_0
-
\frac{i}{\hbar}
\int_0^t ds\,
[\hat H_{\rm int}^{I}(s),\hat\rho^{\rm tot}(s)].
\label{formal}
\end{equation}

Substituting Eq.~\eqref{formal} back into
Eq.~\eqref{VN_B}, one obtains
\begin{align} 
\frac{d}{dt}\hat\rho^{\rm tot}(t)
&=
-\frac{i}{\hbar}
[\hat H_{\rm int}^{I}(t),\hat\rho^{\rm tot}_0]
\nonumber\\
&\quad
-
\frac{1}{\hbar^2}
\int_0^t ds\,
[\hat H_{\rm int}^{I}(t),
[\hat H_{\rm int}^{I}(s),\hat\rho^{\rm tot}(s)]].
\label{VN2_B}
\end{align}

The reduced density matrix of the $a$ subsystem
is defined by
\begin{equation}
\hat\rho_a(t)
=
\Tr_b
\hat\rho^{\rm tot}(t).
\end{equation}

Assuming initially factorized states \cite{PhysRevD.104.083508},
\begin{equation}
\hat\rho^{\rm tot}_0
=
\hat\rho_a(0)\otimes \rho_b,
\end{equation}
and tracing Eq.~\eqref{VN2_B} over the amplified sector
yields
\begin{align}
\frac{d}{dt}\hat\rho_a(t)
&=
-\frac{i}{\hbar}
\Tr_b
[\hat H_{\rm int}^{I}(t),\hat\rho^{\rm tot}_0]
\nonumber\\
&\quad
-
\frac{1}{\hbar^2}
\int_0^t ds\,
\Tr_b
[\hat H_{\rm int}^{I}(t),
[\hat H_{\rm int}^{I}(s),\hat\rho^{\rm tot}(s)]].
\label{red1}
\end{align}

Using Eq.~\eqref{HintI}, the first-order contribution
becomes
\begin{align}
&\Tr_b
[\hat H_{\rm int}^{I}(t),\hat\rho^{\rm tot}_0]
\nonumber\\
&=
-i\hbar\Gamma \Tr_b \left[
\hat a\hat b - \hat a^\dagger \hat b^\dagger,
\hat\rho_a(0)\otimes\rho_b \right] \\
&=
-i\hbar\Gamma
\Big(
\hat a\,\hat\rho_a(0)\,
\Tr_b(\hat b\rho_b) - \hat\rho_a(0)\hat a\,
\Tr_b(\rho_b\hat b) \Big)
\nonumber\\
&\quad
+i\hbar\Gamma \Big(\hat a^\dagger \hat\rho_a(0)\,
\Tr_b(\hat b^\dagger\rho_b) - \hat\rho_a(0)\hat a^\dagger\, \Tr_b(\rho_b\hat b^\dagger) \Big), \nonumber
\end{align} 
where, in the last equality, the commutator has been expanded.

Choosing the $b$ mode initially in the vacuum
state
\begin{equation}
\rho_b
=
|0_b\rangle\langle 0_b|,
\end{equation}
one has
\begin{equation}
\Tr_b(\hat b\rho_b)
=
\Tr_b(\hat b^\dagger\rho_b)
=
0,
\end{equation}
and therefore the first-order contribution vanishes,
\begin{equation}
\Tr_b
[\hat H_{\rm int}^{I}(t),\hat\rho^{\rm tot}_0]
=
0.
\end{equation}

The reduced evolution equation therefore reduces to
\begin{equation}
\frac{d}{dt}\hat\rho_a(t)
=
-\frac{1}{\hbar^2}
\int_0^t ds\,
\Tr_b
[\hat H_{\rm int}^{I}(t),
[\hat H_{\rm int}^{I}(s),\hat\rho^{\rm tot}(s)]].
\label{red2}
\end{equation}

Substituting Eq.~\eqref{HintI} gives
\begin{align}
\frac{d}{dt}\hat\rho_a(t)
&=
-\Gamma^2
\int_0^t ds\,
\Tr_b
\Big[
(\hat a\hat b-\hat a^\dagger\hat b^\dagger),
\nonumber\\
&\qquad\qquad
[(\hat a\hat b-\hat a^\dagger\hat b^\dagger),
\hat\rho^{\rm tot}(s)]
\Big].
\label{double}
\end{align}

Expanding the nested commutator produces terms of the
form
\begin{equation}
\Tr_b(\hat b\hat b\,\rho),
\qquad
\Tr_b(\hat b^\dagger\hat b^\dagger\rho),
\qquad
\Tr_b(\hat b\hat b^\dagger\rho),
\qquad
\Tr_b(\hat b^\dagger\hat b\rho).
\end{equation}

The relevant vacuum correlators are
\begin{align}
\langle 0_b|\hat b\hat b|0_b\rangle
&=
0,
\\
\langle 0_b|\hat b^\dagger\hat b^\dagger|0_b\rangle
&=
0,
\\
\langle 0_b|\hat b^\dagger\hat b|0_b\rangle
&=
0,
\\
\langle 0_b|\hat b\hat b^\dagger|0_b\rangle
&=
1.
\end{align}

The reduced evolution equation (\ref{VN2_B}) can therefore be written
formally as
\begin{align} 
\frac{d}{dt}\hat\rho_a(t)
=
\mathcal L_u\,\hat\rho_a(t)
+
\mathcal K[\hat\rho^{\rm tot}(t)],
\label{eom_bateman}
\end{align}
where
\begin{align}
\mathcal L_u\hat\rho_a(t)
=
-\frac{i}{\hbar}
[\hat H_a,\hat\rho_a(t)],
\label{Lu_bateman}
\end{align}
and
\begin{align}
\mathcal K[\hat\rho^{\rm tot}(t)]
=
-\frac{1}{\hbar^2}
\int_0^t ds\,
\Tr_b
\Big[
\hat H_{\rm int}^{I}(t),
[\hat H_{\rm int}^{I}(s),
\hat\rho^{\rm tot}(s)]
\Big].
\label{Kb}
\end{align}

Although the interaction-picture Hamiltonian remains
time independent in the present case, the kernel
contains an explicit integration over the prior
evolution history of the coupled system through the
time-dependent density operator
$\hat\rho^{\rm tot}(s)$. Consequently, the reduced
subsystem dynamics at time $t$ depends not only on its
instantaneous state but also on its earlier
interaction history, thereby generating an
intrinsically non-Markovian evolution characterized by
a nonlocal memory kernel
\cite{PhysRevD.104.083508}. In contrast to
conventional Markovian dissipation into a large
external reservoir \cite{Lidar:2019qog}, the memory
effects in the present framework originate from the
coherent bidirectional exchange between the damped and
amplified sectors of the globally unitary doubled
Bateman system. From the perspective of the reduced subsystem, the traced-out amplified sector acts as an effective internal environment, yielding an open-system description even though the full Bateman system evolves unitarily \cite{Maity:2024wgl}.


\section{Expectation values of subsystem observables}

The expectation value of a subsystem observable
$\hat O_a$ is defined through the reduced density matrix
as
\begin{equation}
\langle \hat O_a(t)\rangle
=
\Tr_a
\left[
\hat\rho_a(t)\hat O_a
\right].
\label{obs_red}
\end{equation}

The reduced density matrix may formally be written as
\begin{align}
\hat\rho_a(t)
&=
e^{\mathcal L_u t}\hat\rho_a(0)
\nonumber\\
&\quad
+
e^{\mathcal L_u t}
\int_0^t d\tau\,
e^{-\mathcal L_u \tau}
\mathcal K[\hat\rho^{\rm tot}(\tau)],
\label{formal_sol_obs}
\end{align}
where Eq. (\ref{Lu_bateman}) is used.

Substituting Eq.~\eqref{formal_sol_obs} into
Eq.~\eqref{obs_red} gives
\begin{align}
\langle \hat O_a(t)\rangle
&=
\Tr_a
\left[
e^{\mathcal L_u t}
\hat\rho_a(0)\hat O_a
\right]
\nonumber\\
&\quad
+
\Tr_a
\left[
e^{\mathcal L_u t}
\int_0^t d\tau\,
e^{-\mathcal L_u \tau}
\mathcal K[\hat\rho^{\rm tot}(\tau)]
\hat O_a
\right].
\label{obs_general}
\end{align}

Using the unitary propagator
\begin{align}
e^{\mathcal L_u t}A
=
e^{-\frac{i}{\hbar}\hat H_a t}
A
e^{\frac{i}{\hbar}\hat H_a t},
\end{align}
one obtains
\begin{align}
\langle \hat O_a(t)\rangle
&=
\Tr_a
\Big[
e^{-\frac{i}{\hbar}\hat H_a t}
\hat\rho_a(0)
e^{\frac{i}{\hbar}\hat H_a t}
\hat O_a
\Big]
\nonumber\\
&\quad
+
\Tr_a
\Bigg[
e^{-\frac{i}{\hbar}\hat H_a t}
\int_0^t d\tau\,
e^{\frac{i}{\hbar}\hat H_a \tau}
\mathcal K[\hat\rho^{\rm tot}(\tau)]
\nonumber\\
&\qquad\qquad\times
e^{-\frac{i}{\hbar}\hat H_a \tau}
e^{\frac{i}{\hbar}\hat H_a t}
\hat O_a
\Bigg].
\label{obs_explicit}
\end{align}

To illustrate the role of the interaction kernel,
consider the occupation number operator
\begin{equation}
\hat N_a
=
\hat a^\dagger\hat a.
\end{equation}

Its expectation value is
\begin{equation}
\langle N_a(t)\rangle
=
\Tr_a
\left[
\hat\rho_a(t)\hat N_a
\right].
\label{Na_red_full}
\end{equation}

Differentiating Eq.~\eqref{Na_red_full} yields
\begin{align}
\frac{d}{dt}\langle N_a(t)\rangle
&=
\Tr_a
\left[
\frac{d\hat\rho_a(t)}{dt}
\hat N_a
\right].
\end{align}

Substituting the reduced evolution equation gives
\begin{align}
\frac{d}{dt}\langle N_a(t)\rangle
&=
\Tr_a
\left[
\mathcal L_u\hat\rho_a(t)\hat N_a
\right]
\nonumber\\
&\quad
+
\Tr_a
\left[
\mathcal K[\hat\rho^{\rm tot}(t)]
\hat N_a
\right].
\label{Na_dot2}
\end{align}
Since
\begin{equation}
\hat H_a
=
\hbar\Omega_0\hat N_a,
\end{equation}
one has $[\hat N_a,\hat H_a]
= 0$
and therefore
\begin{equation}
\Tr_a
\left[
\mathcal L_u\hat\rho_a\hat N_a
\right]
=
0.
\label{unitary_zero}
\end{equation}

Consequently,
\begin{equation}
\frac{d}{dt}\langle N_a(t)\rangle
=
\Tr_a
\left[
\mathcal K[\hat\rho^{\rm tot}(t)]
\hat N_a
\right].
\label{Na_kernel}
\end{equation}

Substituting the explicit kernel yields
\begin{align} \nonumber
\frac{d}{dt}\langle N_a(t)\rangle
&=\\
-\frac{1}{\hbar^2}
\int_0^t ds
&\Tr_{ab}
\Big[
\hat H_{\rm int}^{I}(t),
[\hat H_{\rm int}^{I}(s),
\hat\rho^{\rm tot}(s)]
\Big]
\hat N_a.
\label{Na_kernel3}
\end{align}

Using cyclicity of the trace,
\begin{align}
\Tr
\Big(
[A,[B,\rho]]N
\Big)
=
\Tr
\Big(
\rho[[N,A],B]
\Big),
\end{align}
one obtains
\begin{align}
\frac{d}{dt}\langle N_a(t)\rangle
&=\\
-\frac{1}{\hbar^2}
\int_0^t ds 
&\Tr_{ab}
\left[
\hat\rho^{\rm tot}(s)
\left[
[\hat N_a,\hat H_{\rm int}^{I}(t)],
\hat H_{\rm int}^{I}(s)
\right]
\right].
\label{Na_kernel4}
\end{align}

Using
\begin{equation}
[\hat N_a,\hat a] = -\hat a, \qquad [\hat N_a,\hat a^\dagger] =  \hat a^\dagger,
\end{equation}
one finds
\begin{align}
[\hat N_a,\hat H_{\rm int}^{I}(t)]
&=
i\hbar\Gamma
\left(
\hat a\hat b
+
\hat a^\dagger\hat b^\dagger
\right).
\label{comm1}
\end{align}

Substituting Eq.~\eqref{comm1} into
Eq.~\eqref{Na_kernel4} yields
\begin{align} \nonumber
\frac{d}{dt}\langle N_a(t)\rangle
&=\\
-\Gamma^2 \int_0^t ds 
&\Tr_{ab}
\Big[
\hat\rho^{\rm tot}(s)
\times
\left[
(\hat a\hat b+\hat a^\dagger\hat b^\dagger),
(\hat a\hat b-\hat a^\dagger\hat b^\dagger)
\right]
\Big].
\label{Na_kernel5}
\end{align}

Equation~(\ref{Na_kernel5}) demonstrates that the evolution of the subsystem occupation number is governed entirely by the interaction kernel generated by the coherent coupling between the observable oscillator and the traced-over sector. Since the free subsystem Hamiltonian commutes with $\hat N_a$, the occupation number would remain constant in the absence of the interaction. Consequently, every change in $\langle N_a(t)\rangle$ originates from information exchange between the two oscillator sectors.

Furthermore, Eq.~(\ref{Na_kernel5}) depends on the full density matrix $\hat\rho^{\rm tot}(s)$ at all previous times, $0\le s\le t$. Therefore the future evolution of the subsystem is not determined solely by its instantaneous state but also by its interaction history. This explicit dependence on past states is the hallmark of non-Markovian dynamics.

To understand the physical origin of this memory, consider the exact dynamics of the full doubled Bateman system. Following the Eqs. (\ref{Apm}) - (\ref{scalinglaw}) and the realted discussion,   
%
%
%
%
%
%
%
we see that  the full doubled system possesses an exact logarithmic-spiral scaling covariance.

Expressed in terms of the original oscillator operators, the observables take the form of $Q_\pm$ expressed in Eq. (\ref{Qpmdecomp}), showing that 
%
%
the operators $ab$, and $a^\dagger b^\dagger$
constitute the essential cross-sector degrees of freedom responsible for generating the self-similar scaling structure of the full doubled system.

After tracing over the $(b)$-sector,
\begin{equation}
\rho_a(t)
=
{\rm Tr}_b
\big[
\rho^{\rm tot}(t)
\big],
\end{equation}
the collective observables $Q_\pm$ are no longer directly accessible because they explicitly involve the traced-out oscillator. Consequently, the reduced subsystem cannot exhibit the exact scaling law satisfied by the full doubled dynamics.

This may be seen directly from the exact solution
\begin{equation}
a(t)
=
e^{-i\Omega_0 t}
\Big[
a\cosh(\Gamma t)
+
b^\dagger\sinh(\Gamma t)
\Big].
\end{equation}

The corresponding number operator becomes
\begin{align}
N_a(t)
=
&
\cosh^2(\Gamma t)\,N_a
+
\sinh^2(\Gamma t)\,(N_b+1)
\nonumber\\
&
+
\cosh(\Gamma t)\sinh(\Gamma t)
\left(
a^\dagger b^\dagger+ab
\right).
\end{align}
Under the discrete translation
\begin{equation}
t\rightarrow t+T,
\end{equation}
the hyperbolic functions mix according to
\begin{equation}
\cosh[\Gamma(t+T)]
=
\cosh(\Gamma t)\cosh(\Gamma T)
+
\sinh(\Gamma t)\sinh(\Gamma T),
\end{equation}
so that
\begin{equation}
N_a(t+T)
\neq
e^{\lambda T}
N_a(t)
\end{equation}
for any constant $\lambda$. Hence the subsystem observable $N_a$ does not inherit the exact scaling covariance of the full doubled system.

The crucial point, however, is that the information responsible for the scaling structure is not destroyed by the partial trace. Rather, it reappears through the history-dependent kernel governing the reduced dynamics. From Eq.~(\ref{Na_kernel5}),
\begin{multline}
\frac{d}{dt}\langle N_a(t)\rangle
=
-\Gamma^2
\int_0^t ds\,
{\rm Tr}_{ab}
\Big[
\rho^{\rm tot}(s)
\\
\times
\big[
(ab+a^\dagger b^\dagger),
(ab-a^\dagger b^\dagger)
\big]
\Big].
\end{multline}

Remarkably, the same operator combinations
$ab$, and $a^\dagger b^\dagger$,
that enter the scaling observables $Q_\pm$ also appear explicitly inside the memory integral. This shows that the degrees of freedom responsible for the scaling covariance of the full doubled system are not eliminated by the partial trace. Rather, their influence survives indirectly through the history-dependent kernel governing the reduced dynamics.

In this way, the information associated with the self-similar amplified and damped branches is transferred from  
collective quantities into the temporal correlations controlling the non-Markovian evolution of the subsystem.

The partial trace therefore performs two distinct operations simultaneously. First, it removes direct access to the collective observables $Q_\pm$, thereby destroying the observable scaling covariance at the subsystem level. Second, it preserves the cross-sector correlations generated by the same operators $ab$ and $a^\dagger b^\dagger$, which continue to influence the subsystem through the integral over past times.

In this sense, the non-Markovian memory kernel may be viewed as the residual dynamical manifestation of the self-similar amplified and damped branches of the full doubled Bateman system. The exact scaling law is no longer visible in subsystem observables, but the operator content responsible for that scaling survives through the history dependence of the reduced evolution.

\section*{Acknowledgments}
PN acknowledges support from the National Institute for Theoretical and Computational Sciences (NITheCS) through the Rector’s Postdoctoral Fellowship Programme (RPFP). We are grateful to Maxim Chernodub and Maciej Malinowski for their insightful, critical, and conceptually stimulating comments on the first version of this manuscript, which helped improve the present work. PN also thanks Prof.~Antonio Capolupo of the University of Salerno for the invitation to the conference ``Quantum Universe 2025,'' held in Avellino, Italy, in November 2025, where this work was initiated. He further thanks the organizers for providing a stimulating and enjoyable environment that fostered fruitful scientific discussions and contributed to the development of this work.

\bibliographystyle{apsrev4-1}
\bibliography{gw_phases}

\end{document}